\documentclass{elsart}

\usepackage{graphicx}
\usepackage{amssymb}

\begin{document}

\begin{frontmatter}

\title{Vertical transmission of culture and the distribution of
family names}

\author[Z]{Dami\'an H. Zanette},
\ead{zanette@cab.cnea.gov.ar}
\author[M]{Susanna C. Manrubia}
\ead{manrubia@ mpikg-golm.mpg.de}

\address[Z]{Consejo Nacional de Investigaciones Cient\'{\i}ficas y
T\'ecnicas, Centro At\'omico Bariloche and Instituto Balseiro, 8400
Bariloche, R\'{\i}o Negro, Argentina}

\address[M]{Max Planck Institute of Colloids and Interfaces, Theory
Division, D-14424 Potsdam, Germany}

\begin{abstract}
A  stochastic  model  for  the  evolution  of  a growing population is
proposed, in order to explain empirical power-law distributions in the
frequency  of  family  names  as  a  function  of  the  family   size.
Preliminary  results  show  that  the  predicted exponents are in good
agreement with real data.  The evolution of family-name  distributions
is  discussed  in  the  frame  of  vertical  transmission  of cultural
features.
\end{abstract}

\end{frontmatter}

\section{Introduction}

\begin{figure}[h]
\centering
\resizebox{\columnwidth}{!}{\includegraphics{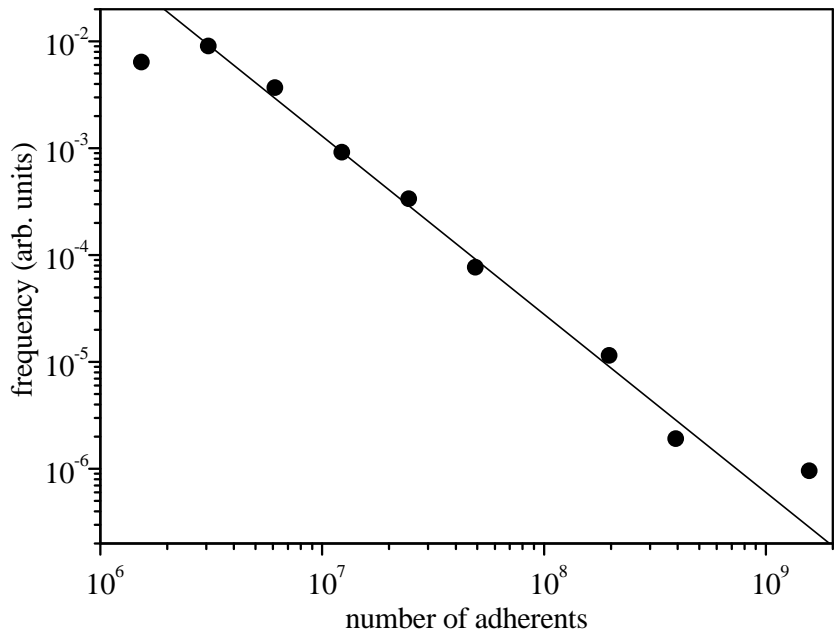}}
\caption{Frequency of religious groups as a function of the number  of
adherents,  in  arbitrary  units  (source:   www.adherents.com).   The
straight line has slope $-5/3 \approx -1.67$.}
\label{f1}
\end{figure}

The  fascinating  complexity  of  social  phenomena  is   increasingly
attracting the attention of physicists.  We find in the techniques  of
Statistical Physics  an ideal  tool for  the study  of models  of such
phenomena,   where   complex    ``macroscopic''   behaviour    emerges
spontaneously as the consequence of relatively simple  ``microscopic''
dynamical rules.   During the  last decade,  in fact,  much work along
those lines has been devoted to the study of statistical properties of
dynamical processes  in economics  \cite{eco1,eco2}. Other  key social
processes---such as the dynamics of cultural features---have  received
relatively less attention,  in spite of  the fact that  empirical data
call  for  the  kind  of  approach  already  employed  with economical
systems.   Consider,  for  instance,  the  size  distribution of large
religious groups, shown in Fig.   \ref{f1}.  A well defined  power-law
decay, spanning more that two orders of magnitude, is apparent.  These
power-law distributions are indeed a  main clue to complexity in  real
and model systems \cite{eco2}.

The  spatiotemporal  dynamics  of  culture  is  driven by geographical
dissemination of cultural features and by their transmission from  old
to  new  generations.   Axelrod  \cite{Axelrod}  has proposed a simple
model of  culture dissemination  that captures  its basic  mechanisms.
Cultural features can spread  by interaction between individuals,  but
some preexistent cultural agreement is necessary for such  interaction
to take place.  These  mechanisms are able to explain  the maintenance
of a certain level of cultural diversity.  Meanwhile, vertical culture
transmission---along the  genealogical line,  from ancestors  to their
descendents---is governed by the influence of cultural features in the
formation of couples, and by  the influence of each parent's  features
in determining those of the offspring \cite{CS1}.  Cavalli-Sforza  and
coworkers have  modeled and  studied different  situations of vertical
culture  transmission,  with  special   emphasis  on  the  effect   of
stochastic external agents \cite{CS2}.

\begin{figure}[h]
\centering
\resizebox{\columnwidth}{!}{\includegraphics{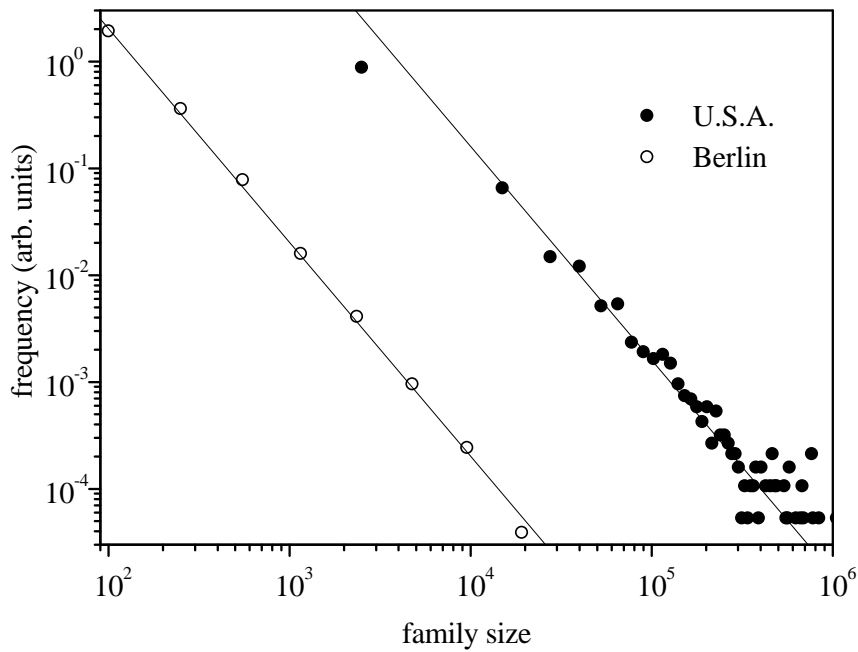}}
\caption{Frequency of family names as  a function of the family  size,
in arbitrary  units. The  United States  data is  extrapolated from  a
sample taken during  the 1990 census  (source:  www.census.gov).   The
Berlin data corresponds to family names beginning by A, taken from the
1996 phonebook. Family  sizes in the  Berlin data are  multiplied by a
factor $10^2$ for convenience in displaying.  The straight lines  have
slope $-2$.}
\label{f2}
\end{figure}

An extreme case  of vertical transmission  of a ``cultural''  feature,
which can be used  as a benchmark for  models of culture dynamics,  is
that of family names. An  individual's family name is (in  most cases,
at  least)  inherited  from  the  father  and, therefore, its possible
influence in the formation of the parents' couple is irrelevant to its
transmission.   Moreover, creation  and mutation  of family  names are
strongly restricted to specific  historical periods and places.   Most
of the time, such changes are  extremely rare.  The history of  family
names is, in fact, quite complex \cite{PlS}.  In Europe, for instance,
different  groups  of  family  names (patronymic-like, toponymic-like,
etc.)  originated  at  different  times---typically, during the Middle
Ages---and  mutations  became  important particularly during the large
migration waves within  Europe and towards  the Americas.   New family
names appeared also as a consequence of immigration.  In spite of this
eventful  history,  current  distributions  of  family  names  exhibit
striking regularities. Figure  \ref{f2} shows family-name  frequencies
as a function of the family size---i.e., of the number of  individuals
bearing a  given family  name---for the  United States  and a  part of
Berlin, in  recent  times.   Both  data  show a well defined power-law
dependence,  with  an  exponent  close  to  $-2$. Analogous  data have
recently been  reported for  Japanese family  names \cite{jap},  which
exhibit  power-law  distributions  with  smaller  exponents  ($\approx
-1.75$).

In this paper, we consider a model for a growing population where each
individual  can  inherit  cultural  features  from  its  parents.   In
particular, we analyze  the case of  transmission of the  family name,
and  study  its  distribution  as  a  function of the family size. The
parameters relevant to  the model are  the relative birth rate and the
mortality, which control  the population growth, and  the creation
rate of  family names.   Our preliminary  results show  that the model
satisfactorily reproduces the  power laws observed  in real data,  for
wide ranges of the parameters.

\section{The model}

We introduce in the following a variation of the mechanism proposed by
Simon \cite{Simon}  to explain  the occurrence  of power  laws in  the
frequency  distribution   of  words   and  city   sizes  (Zipf's   law
\cite{Zipf}), among other instances.  In our model, evolution proceeds
by discrete steps.  At a given step $s$, the $P(s)$ individuals in the
population are divided into groups---the families.  Within each group,
all the individuals  share the same  family name.   At each step,  two
mechanisms act.  (i) A new individual is introduced in the population,
representing a birth event.  With probability $\alpha$ the newborn  is
assigned a new family name, not previously present in the  population.
With  the   complementary  probability,   $1-\alpha$,  a   preexistent
individual is chosen at random to become the newborn's father, and its
family name is given to the  newborn. Thus, a specific family name  is
assigned with a probability  proportional to the corresponding  family
size. (ii) An individual is chosen at random from the whole population
and, with probability $\mu$, it is eliminated. This represents a death
event. Note that if the dead  was the only individual with its  family
name, this specific family name disappears from the population.

The evolution of  the population is  controled by the  parameter $\mu$
which, as we show  below, is a direct  measure of the mortality  rate.
The  distribution  of  family  names  varies  due  to  the  effect  of
family-name  creation  and  mutation,  measured  by  $\alpha$,  and of
mortality.   Since during  the evolution  the total  population $P(s)$
changes, the time  interval $\delta t(s)$  to be associated  with each
evolution step should also change, as $\delta t(s) = 1/\nu P(s)$.  The
frequency $\nu$,  whose value  is in  principle arbitrary,  fixes time
units.  The variation of the  population at each step is, on average,
$\delta P(s)=1-\mu$. Consequently, the ``macrosopic'' equation for the
time evolution of the population reads
\begin{equation}    \label{dP/dt}
\frac{dP}{dt} \approx \frac{\delta P}{\delta t}=
\nu (1-\mu) P.
\end{equation}
Identifying $\nu$ with  the birth rate  per individual and  unit time,
the  product  $\nu  \mu$  is  the  corresponding  mortality  rate.  In
average, thus, the population grows exponentially in time.

Note that, since  an individual's family  name is here  supposed to be
inherited from the  father, the model  describes the evolution  of the
male population only. However, the same mechanism can be reinterpreted
assuming that the family name is transmitted with the same probability
by  either  parent.  In  this  case,  the  model encompasses the whole
population and no  sex distinction occurs.   The real situation  is in
fact intermediate  between these  two limiting  cases. We  also stress
that in the present model  individuals are ageless, in the  sense that
neither the probability of becoming father of a newborn nor the  death
probability depend  on the  individual's age.   As a  consequence, the
probability  $p(m)$  that  an  individual  has $m$ children during its
whole life is exponential, $p(m)=\mu  (1+\mu)^{-m-1}$.  This is to  be
compared  with  the  Poissonian  probability  of  real, age-structured
populations \cite{Poiss}.

Below, we consider a class of initial conditions where the  population
is divided into $N_0$ families, with $i_0$ individuals in each family.
We denote such an initial condition as $(N_0,i_0)$. The  corresponding
initial population is $P(0)=N_0 i_0$.

\subsection{Simon's model: $\mu = 0$}

Neglecting mortality (that is with  $\mu = 0$), our system  reduces to
the model introduced by Simon  to explain Zipf's law \cite{Simon}.  In
this  case,  the  evolution   of  the  population  is   deterministic,
$P(s)=P(0)+s$, since exactly one individual is added to the population
at each  step.   Under these  conditions, it  is possible  to write an
evolution equation for  the average number  of families $n_i(s)$  with
exactly $i$ individuals at step $s$. We have
\begin{equation}   \label{Simon2}
n_i(s+1)=n_i(s)+\frac{1-\alpha}{P(s)}\left[
(i-1) n_{i-1}(s)-in_i(s) \right],
\end{equation}
for $i>1$, and
\begin{equation}    \label{Simon3}
n_1(s+1)=n_1(s)+\alpha-\frac{1-\alpha}{P(s)}n_1 (s).
\end{equation}
Simon has shown that, under fairly general conditions, these equations
predict a long-time distribution with a power-law decay
\begin{equation}    \label{Simon4}
n_i \propto i^{-1-1/(1-\alpha)}
\end{equation}
for  moderately  large  values  of  $i$  ($1\ll  i  \ll  N_0+s$). This
power-law  distribution   is  to   be  ascribed   to  the   stochastic
multiplicative  nature  of  family  growth,  which  involves  a growth
probability proportional to the family size. In the limit $\alpha  \to
0$ the exponent in  Eq.  (\ref{Simon4}) equals  $-2$.  Note that  this
limit is relevant to our problem, since the probability of creation or
mutation of a family name per individual is expected to be very small.
The exponent,  in fact,  agrees with  the empirical  data presented in
Fig. \ref{f2}.

\begin{figure}[h]
\centering
\resizebox{\columnwidth}{!}{\includegraphics{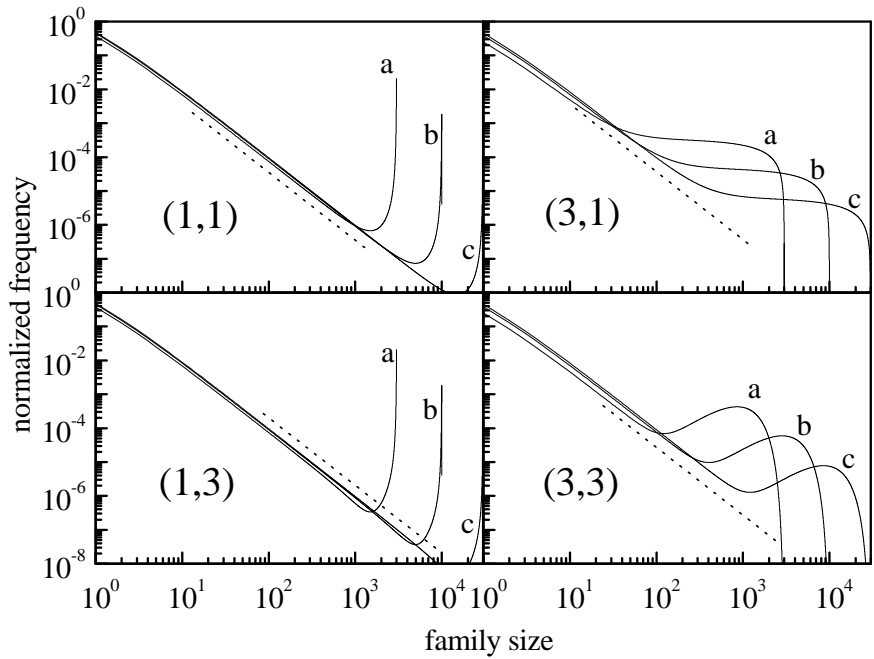}}
\caption{Family-name frequency as a function of the family size, given
by Eqs.  (\ref{Simon2}) and  (\ref{Simon3}) with  $\alpha=10^{-3}$, at
three  evolution  stages:  (a)  $s=3\times  10^3$,  (b)  $s=10^4$, (c)
$s=3\times 10^4$.   For convenience  in displaying,  the distributions
have  been  normalized.  The  numbers  in  brackets  give  the initial
condition for each  case (see text).   The dotted straight  lines have
slope $-2$.}
\label{f3}
\end{figure}

We  point  out  that  transient  effects  strongly  depend  on initial
conditions.   Figure  \ref{f3}  shows  the  (normalized)  distribution
$n_i(s)$ calculated  from Eqs.   (\ref{Simon2}) and  (\ref{Simon3}) at
several  evolution  stages,  for  different  initial  conditions   and
$\alpha=10^{-3}$.  For intermediate  values of $i$ the  development of
the power-law  decay with  exponent close  to $-2$  is apparent in all
cases.  However, the behaviour  of the distribution for larger  values
of $i$ varies noticeably with the initial condition.

Equations  (\ref{Simon2})  and  (\ref{Simon3})  imply  that  the total
number  of  family  names  in  the  population, given by $N(s)= \sum_i
n_i(s)$, grows  in average  as $N(s)=N_0+\alpha  s$. As  a function of
time, thus,  the number  of family  names increases  exponentially, as
$N(t)  =N_0  \exp(\alpha  t)$,  as  expected  for a population without
mortality  where  family  names  are  created  at  rate  $\alpha$.  In
contrast, in real populations at  present times, the number of  family
names is known to decrease \cite{PlS}.

\subsection{Effects of mortality: $\mu \neq 0$}

With  $\mu  \neq  0$,  the  growth  of  the  total  population  $P(s)$
fluctuates stochastically, depending on the occurrence of death events
at each evolution  step. Consequently, a  formulation for the  average
evolution of $n_i(s)$ in terms of a deterministic equation of the form
of Eqs. (\ref{Simon2}) and (\ref{Simon3}) turns out to be inconsistent.
These equation can however be adapted in a way suitable for  numerical
calculation  to   the  case   where  the   population  growth   is not
deterministic, in  the following  form. First,  for a  given value  of
$P(s)$  at  step  $s$,  the  functions  in the right-hand side of Eqs.
(\ref{Simon2}) and  (\ref{Simon3}) are  applied to  $n_i(s)$ to obtain
intermediate  values  $n_i'(s)$.   Then,  with  probability  $\mu$, we
calculate
\begin{equation}    \label{mu1}
n_i(s+1) = \frac{1}{P(s)+1} \left[ (i+1) n_{i+1}'(s)-in_i'(s)
\right]
\end{equation}
for all $i=1,2,\dots$. Since in this case both birth and death  events
have taken  place, $P(s+1)=P(s)$.  With the complementary probability,
$1-\mu$, we put $n_i(s+1)=n_i'(s)$ for all $i$, and $P(s+1)=P(s)+1$.

\begin{figure}[h]
\centering
\resizebox{\columnwidth}{!}{\includegraphics{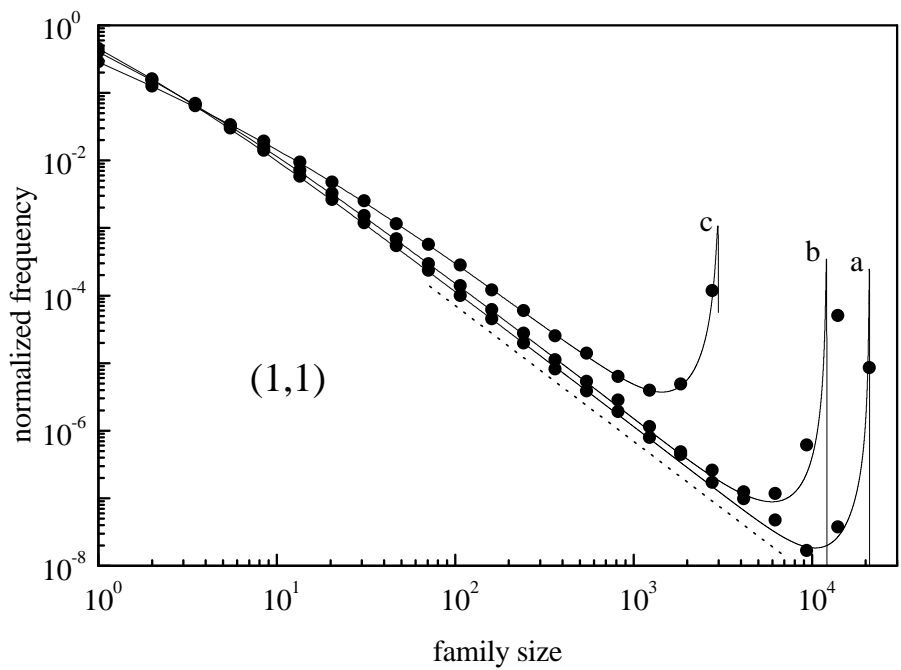}}
\caption{Normalized family-name frequency as a function of the  family
size  calculated   from  Eqs.   (\ref{Simon2}),  (\ref{Simon3})    and
(\ref{mu1}) at $s=3\times 10^4$, for $\alpha=10^{-3}$ and three values
of $\mu$ (a) $\mu=0.3$, (b)  $\mu=0.6$, (c) $\mu=0.9$. Dots stand  for
the results of numerical simulations of the model, averaged over  $10^4$
realizations. The dotted straight line has slope $-2$.}
\label{f4}
\end{figure}

Heuristic arguments---not reproduced  here---indicate that, under  the
conditions  used  to  derive  Eq.  (\ref{Simon4}), the above algorithm
should give rise to distributions  with a well defined power-law decay
for moderately large family sizes, of the form
\begin{equation}    \label{mu2}
n_i \propto i^{-1-(1+\mu)/(1+\mu-\alpha)}.
\end{equation}
Quite remarkably,  in the  relevant limit  $\alpha\to 0$  the exponent
becomes  independent  of  $\mu$,  and  reduces  again  to  $-2$.   For
sufficiently low $\alpha$, thus,  mortality is not expected  to affect
the power-law exponent  which, as we  have seen, is  in agreement with
empirical data. This has  been verified through numerical  calculation
of $n_i(s)$ with the above algorithm, as illustrated in Fig.  \ref{f4}
for  the initial condition $(1,1)$ and three values of $\mu$.

The  algorithm  combining  Eqs.   (\ref{Simon2}),  (\ref{Simon3}), and
(\ref{mu1}) mixes the deterministic average evolution of $n_i(s)$ with
the  stochastic  variation  of  the  population,  due  to random death
events. This combination  involves, thus, a  statistical approximation
which must be  tested by means  of numerical simulations  of the fully
stochastic model.   Results of  such simulations,  averaged over  $10^4$
realizations  for  each  value  of  $\mu$,  are  shown as dots in Fig.
\ref{f4}. We find very good agreement between both methods.

As for  the number  of different  family names,  $N(t)$, we have found
that, for moderate values of $\mu$ and at sufficiently long times,  it
increases exponentially. As expected, the growth rate depends on  both
$\alpha$ and $\mu$. There is however an initial transient during which
the evolution is not exponential and, in fact, $N(t)$ can  temporarily
decrease. Decay of the number of family names for long times seems  to
be restricted to  very high death  probability, $\mu \approx  1$. Note
that  these  are  precisely  the  values  expected for $\mu$ in modern
developed  societies,  where  birth  and  death  rates are practically
identical.

\section{Discussion}

The present variant of Simon's model provides a plausible  description
of a growing population, as  far as the assumption of  age-independent
fertility  and  mortality  is  admitted.   The numerical resolution of
averaged evolution equations and  numerical simulations show that  our
model successfully reproduces the exponent of power-law  distributions
observed in the frequency of family names as a function of the  family
size.   Specifically, the  exponent close  to $-2$  found in empirical
data for family names from the United States and Berlin is  reproduced
in the  limit of  very small  creation and  mutation rates  and a wide
variety of mortality rates.

For other creation and  mutation rates,  the predicted  exponents are,
in absolute value, larger  than above  [cf. Eqs.  (\ref{Simon4}) and
(\ref{mu2})].  This contrasts with  the exponents found  for modern
Japanese  family names, close to $-1.75$ \cite{jap}. We argue that
this is an effect of transients  which,  in  this  case,  are  still
acting. In fact, most Japanese family names are relatively recent, as
they appeared some 120 years ago \cite{jap}. Curve (c) in Fig.
\ref{f4}, for instance,  shows clearly  that  transient  distributions
could  be  assigned   smaller spurious power-law exponents.  Note
however that a detailed evaluation of  transient  effects  requires  a
careful identification of initial conditions which, as a result of the
complex history of family  names, could be a hard task in any real
situation.

A quantitative comparison of the predictions of the present model with
real  data---not  presented  at  this preliminary level---will require
considering  populations  of  several  million  individuals  (cf. Fig.
\ref{f2}). Since  extensive numerical  simulations of  systems of such
sizes could become computationally too expensive, it will be useful to
analyze in detail the scaling properties of our model. In  particular,
the  attention  will  focus  on  the  dependence  of  the  duration of
transients, both in  the frequency and  in the total  number of family
names, on the initial population and its distribution in families,  as
well  as  on  the  probabilities  $\alpha$  and  $\mu$.    Considering
long-term  variations  of  these   probabilities  is  also  in   close
connection with the comparison of our results with empirical data.  In
fact, for a  modern developed population,  in Europe for  instance, we
can distinguish at  least two well  differentiated stages.   When most
European  family  names  appeared,  some  centuries  ago,  the   total
population was  increasing more  or less  steadily. This  stage, thus,
corresponds to relatively large values of $\alpha$ and moderate values
of $\mu$. In modern times, on the contrary, new family names appear at
an  extremely  low  rate---in  fact,  their  total  number   decreases
\cite{PlS}---and  the   total  European   population  is   practically
constant, so that $\alpha \approx 0$ and $\mu \approx 1$.

The adaptation of the present model  to the study of the evolution  of
other  cultural  features  requires   the addition of  two  main   new
ingredients. First, a new parameter  must be introduced to define  the
probability that  a given  cultural feature  is inherited  from either
parent \cite{CS1,CS2}.  Second, it is necessary to specify the  effect
of  that  feature  in  the  formation  of the parents' couple, and the
mechanism  by  which  couples  are  effectively  formed.   This latter
process  has  been  classically  proposed  as  an optimization problem
\cite{coup}. In the frame of our system, it would require a much  more
realistic approach if any connection with actual populations is to  be
established.

\section*{Acknowledgement}

We thank G. Abramson for his critical reading of the manuscript.

\end{document}